# Synthesis and Characterization of $Na_{0.3}RhO_2 \cdot 0.6H_2O$ - a Semiconductor with a Weak Ferromagnetic Component.

S. Park[1], K. Kang[2], W. Si[1], W.-S. Yoon[2], Y. Lee[1], A. R. Moodenbaugh[2], L. H. Lewis[2] & T. Vogt[1]

[1]Physics Department, Brookhaven National Laboratory, Upton, NY 11973-5000
[2]Materials Science Department, Brookhaven National Laboratory, Upton, NY 11973-5000

*We have prepared the oxyhydrate $Na_{0.3}RhO_2 \cdot 0.6H_2O$ by extracting $Na^+$ cations from $NaRhO_2$ and intercalating water molecules using an aqueous solution of $Na_2S_2O_8$. Rietveld refinement, thermogravimetric analysis (TGA), and energy-dispersive x-ray analysis (EDX) reveal that a non-stoichiometric $Na_{0.3}(H_2O)_{0.6}$ network separates layers of edge-sharing $RhO_6$ octahedra containing $Rh^{3+}$($4d^6$, S=0) and $Rh^{4+}$ ($4d^5$, S=1/2). The resistivities of $NaRhO_2$ and $Na_{0.3}RhO_2 \cdot 0.6H_2O$ (T < 300) reveal insulating and semi-conducting behavior with activation gaps of 134 meV and 7.8 meV, respectively. Both $Na_{0.3}RhO_2 \cdot 0.6H_2O$ and $NaRhO_2$ show paramagnetism at room temperature, however, the sodium-deficient sample exhibits simultaneously a weak but experimentally reproducible ferromagnetic component. Both samples exhibit a temperature-independent Pauli paramagnetism, for $NaRhO_2$ at T > 50 K and for $Na_{0.3}RhO_2 \cdot 0.6H_2O$ at T > 25 K. The relative magnitudes of the temperature-independent magnetic susceptibilities, that of the oxide sample being half that of the oxyhydrate, is consistent with a higher density of thermally accessible electron states at the Fermi level in the hydrated sample. At low temperatures the magnetic moments rise sharply, providing evidence of localized and weakl -ordered electronic spins with effective moment per formula unit values of $2.0 \times 10^{-1}$ $\mu_B$ for $NaRhO_2$ and $0.8 \times 10^{-1}$ $\mu_B$ for $Na_{0.3}RhO_2 \cdot 0.6H_2O$*

E-mail address: tvogt@bnl.gov

## 1. Introduction

Manipulation of the structure of layered oxides can result in novel electronic properties. The de-intercalation and hydration of the layered sodium cobalt oxide $Na_xCoO_2$ through an oxidation and subsequent hydration reaction results in the bilayer hydrate $Na_{0.3}CoO_2\cdot1.4H_2O$, which is superconducting below 4~5 K [1]. An alternative synthesis route to make the superconductive $Na_{0.3}CoO_2\cdot1.4H_2O$ using aqueous $Na_2S_2O_8$ instead of $Br_2$ in acetonitrile and subsequently re-hydrating was introduced in a previous report [2]. The extraction of alkali-metal ions ($A^+$) is accomplished by persulfate-based oxidants, which oxidize the metal M in the $MO_2$ layer:

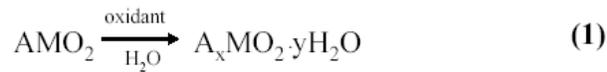

$$AMO_2 \xrightarrow[H_2O]{oxidant} A_xMO_2\cdot yH_2O \qquad (1)$$

In this paper, we report on the synthesis, structure, magnetic, and transport properties of the layered compound $Na_{0.3}RhO_2\cdot0.6H_2O$. This material is isostructural to the monolayer hydrate $Na_{0.3}CoO_2\cdot0.6H_2O$ [3,4]. In $Na_{0.3}RhO_2\cdot0.6H_2O$, $Rh^{3+}$ ($4d^6$) and $Rh^{4+}$ ($4d^5$) ions are both present in octahedral coordination. Due to the relatively large crystal field splitting observed in the second transition-metal series, $Rh^{3+}$ and $Rh^{4+}$ in octahedral sites give rise to spin states $S = 0$ and $\frac{1}{2}$, respectively [5,6,7]. The lower on-site Stoner and Coulomb parameters and the larger band widths in 4d and 5d materials generally suppress magnetism in semiconducting compounds. We show that in $Na_{0.3}RhO_2\cdot0.6H_2O$ a very weak ferromagnetic component and semiconducting behavior coexist at room temperature.

## 2. Experimental

The parent compound $NaRhO_2$ was prepared by heating a 10 mol% excess of $Na_2CO_3$ (Alfa, 99.5%) with $Rh_2O_3$ (Alfa, 99.9%) at 800 °C for 8 h in an $O_2$ (g) atmosphere. The layered compound $Na_{0.3}RhO_2\cdot0.6H_2O$ was then made by stirring $NaRhO_2$ (molar ratio 4:1) in an aqueous solution of $Na_2S_2O_8$ (pH ~ 10.5) for 22 h in a beaker covered with Parafilm[TM]. The Na (x ~ 0.3) content in $Na_xRhO_2 \cdot yH_2O$ was determined by energy dispersive X-ray analysis (EDX) using $NaRhO_2$ as a standard. The water content (y ~ 0.6) was determined using thermogravimetric analysis (TGA, TA Instruments 2960). High-resolution synchrotron X-ray powder diffraction data from $Na_{0.3}RhO_2 \cdot 0.6H_2O$ were measured at beam line X7A at the National Synchrotron Light Source at Brookhaven National Laboratory employing a gas-proportional position-sensitive detector (PSD), using a gated Kr-escape peak [8]. Conductivity measurements were performed on cold-pressed pellets using a standard four-probe measurement setup with a current of 100 μA for the conducting sample and 0.1 μA for the insulating sample. Magnetization measurements were carried out using a SQUID magnetometer in the temperature range 4.5 K -300 K at applied fields 50 kOe ≤ H ≤ -50 kOe. Care was exercised to avoid introduction of magnetic impurities during sample mounting and manipulation and the magnetic results were reproduced several times. To examine the electronic state and the atomic environment of the Rh ion in the compounds, Rh K-edge X-ray absorption spectroscopy (XAS) experiments, in particular XANES (X-ray absorption near-edge electron spectroscopy), were performed on beam line X11A of the National Synchrotron Light Source. The data were analyzed to obtain the extended x-ray absorption fine structure (EXAFS) function $\chi(k)$ using established procedures. The data were then Fourier-transformed from k-space to real space and remain uncorrected for the photoelectron phase shifts; thus determined distances are ~ 0.3-0.4 Å shorter than the actual distances.

3. Results and discussion

The oxidation and water intercalation of $NaRhO_2$ using $Na_2S_2O_8$ in an aqueous solution resulted in significant Bragg peak shifts of the (*00l)* X-ray reflections towards larger d-spacings, while preserving the rhombohedral symmetry (R3-m). We also note that the intercalated $Na_{0.3}RhO_2 \cdot yH_2O$ powder exhibits a dark brown color compared to the orange-yellowish color of the $NaRhO_2$ powder. The TGA data of $Na_{0.3}RhO_2 \cdot yH_2O$ obtained using heating rates of 0.25 ºC/min in flowing Ar revealed a weight loss of about 7 % occurring at 100 ºC. This result corresponds to a loss of 0.6 water per formula unit.

The measured lattice parameters are $a$ = 3.0542(1) Å, $c$ = 20.8560(9) Å for Na$_{0.3}$RhO$_2$·0.6H$_2$O and $a$ = 3.0971(4) Å, $c$ = 15.5275(34) Å for NaRhO$_2$ (ICSD (Inorganic Crystal Structure Database) #66280), respectively. The oxidation of Rh$^{3+}$ to Rh$^{4+}$ caused by the de-intercalation of sodium ions in NaRhO$_2$ results in a contraction of the $a$-axis and a decrease of the average Rh-O bond distances in the RhO$_6$ octahedra. The water intercalation leads to an increase of the $c$-axis parameter. The structural refinement employing the space group R-3m used an expanded layer model of the sodium rhodate structure with the Rh and O atoms positioned at (0,0,0) and (0,0,0.23), respectively, as a starting model. All refinements were carried out using the Rietveld method and the GSAS package [9,10,11]. Difference Fourier maps were generated that allowed location of the other atoms within the structure. In the final refinement, occupancies for the water and sodium cations were fixed to the stoichiometry values obtained from the TGA/EDX measurements. Constraints were used for the isotropic displacement parameters. The final profile fit is depicted in Fig. 1. The parameters of the refined structural model are summarized in Table 1 and selected interatomic distances are listed in Table 2. The structure of Na$_{0.3}$RhO$_2$·0.6H$_2$O consists of layers of rhodium oxide octahedra with water molecules and sodium cations intercalated between them as shown in Fig. 1 (inset). The water molecules, located at the $m$ site with $z \approx 0$, surround the sodium cations at the $-3m$ site, thereby forming a single layer between the RhO$_2$ layers. As a result, the interlayer spacing expands by ca. 34% compared to that of anhydrous sodium rhodate (Fig. 1 (inset)). It is interesting to note that the deintercalation of sodium and the insertion of H$_2$O leads to a shift of the RhO$_2$ layers with respect to their neighboring layers. In sodium rhodate hydrate, the triangular oxygen lattices in the neighboring RhO$_2$ octahedral layers are stacked along their edges, whereas in anhydrous sodium rhodate, they are staggered (Fig. 1 (inset)). This allows the sodium-oxygen coordination (Na–O = 2.529(3) Å) to be retained in the expanded layers. Interatomic distances involving the water oxygen atoms also suggest hydrogen bonding within the Na$_{0.3}$(H$_2$O)$_{0.6}$ layer (Table 2). Further details concerning the structure however, need to be elucidated by neutron powder diffraction. In contrast to the cobalt analogue, Na$_{0.3}$RhO$_2$·0.6H$_2$O is stable when exposed to humid air and appears to be the only stable hydrate in the Na-Rh-O system. Attempts to make bilayer hydrates were unsuccessful.

Electrical resistivity measurements of $NaRhO_2$ and $Na_{0.3}RhO_2 \cdot 0.6H_2O$ are shown in Fig. 2. In $Na_{0.3}RhO_2 \cdot 0.6H_2O$ the electrical resistivity increases with decreasing temperature. At room temperature the magnitude of the $NaRhO_2$ resistivity exceeds that of $Na_{0.3}RhO_2 \cdot 0.6H_2O$ by a factor of $10^4$. The temperature dependence of the resistivity shows thermally activated behavior in both samples at temperatures above T ~ 150 K, but with very different activation energies. By fitting the data to an Arrhenius law *ρ=ρ₀ exp(Δ/kT)*, we extract activation energies of 134 meV and 7.8 meV for $NaRhO_2$ and $Na_{0.3}RhO_2 \cdot 0.6H_2O$, respectively. The activation energy gap is thus significantly reduced in the oxidized and intercalated material. The resistivity of $NaRhO_2$ at low temperatures (T < 50 K) deviates from the activated behavior, gradually adopting a variable range hopping (VRH) behavior.

The positive shift in the edge-position of the Rh K-edge XANES spectrum indicates that, as expected, the average valence of Rh increases (Fig. 3(a)) upon modification of the $NaRhO_2$ compound to form $Na_{0.3}RhO_2 \cdot 0.6H_2O$. The first and second peaks in Fig. 3(b) are due to Rh-O correlations and Rh-Rh interactions, respectively. Peaks both at ~4.9 Å and ~5.8 Å are due to Rh-Rh interactions. The Fourier transformation of the Rh K-edge EXAFS spectrum for $NaRhO_2$ shows a peak intensity at ~5.8 Å that is much larger than that of the peak at ~ 4.9 Å and is attributed to an enhancement by multiple scattering of the EXAFS photoelectron. This peak is seen in several transition-metal oxides with a hexagonal layered structure and arises from a three-body correlation (often referred to as a "focusing effect") involving the coherent linear chains of edge-sharing transition-metal octahedra [12,13]. The intensity of this peak is highest when the dihedral angle between the metals is 180°. Deviations from this colinearity, due to disorder-induced local distortions or broken chains caused by vacancies, lead to a drastic reduction in the peak height. The peak at R ~5.8 Å is much weaker in the $Na_{0.3}RhO_2 \cdot 0.6H_2O$ data, indicating the presence of a significantly higher degree of local disorder in the $RhO_2$ sheets of this layered compound.

The magnetic data collected as a function of temperature (Fig. 4(a)) indicates that both compounds show small temperature-independent average magnetic susceptibilities ($\chi_{av}$) at higher temperatures. Both susceptibilities increase abruptly at low temperatures. The susceptibility increases occur in the parent $NaRhO_2$ compound for T < 50 K from $\chi_{av}$

= 0.92 x $10^{-8}$ m$^3$/kg (0.73 x $10^{-6}$ emu/g-Oe) and in the layered Na$_{0.3}$RhO$_2$·0.6H$_2$O compound for T < 25 K from $\chi_{av}$ = 2.30 x $10^{-8}$ m$^3$/kg (1.83 x $10^{-6}$ emu/g-Oe). Overall, the magnetization trends are consistent with Pauli paramagnetic behavior indicating the existence of thermally accessible energy states at the Fermi level at higher temperatures followed by the condensation of localized moments at low temperatures. The magnitude of the Pauli paramagnetic susceptibility in the stoichiometric compound is smaller by a factor of two than that of the sodium-deficient compound; this is consistent with the larger relative conductivity noted in that sample. It is surmised that the condensation of localized magnetic moments at low temperature is associated with the change of transport behavior from activated to hopping-type electron conductivity. Below 50 K the susceptibility was further analyzed by fitting the data to the equation

$$\chi_{tot} = (C/(T+\theta)) + \chi_P \qquad (2)$$

where $C$ is the Curie constant per gram, $\theta$ is the paramagnetic Curie temperature, and $\chi_P$ is the temperature-independent Pauli paramagnetism. The parameter $\theta = \lambda C$, where $\lambda$ is the molecular field constant within the Weiss theory of ferromagnetism. Within experimental error in both samples, the ($\theta$) ~ 0 K indicates of very weak or zero long-range collective order amidst the localized spins.

The Curie constant per gram $C$ is given by

$$C = \frac{N\mu_{eff}^2}{3Mk_B}, \qquad (3)$$

where $N$ is Avogadro's number, $M$ is the formula unit weight, $k_B$ the Boltzman constant, and $\mu_{eff}$ the effective moment. Hypothesizing that the compounds' magnetic susceptibility signal consists of a thermally assisted electronic contribution (the Pauli susceptibility) superimposed by a localized Curie-Weiss susceptibility indicates that the sodium-deficient sample possesses a much lower Curie constant (0.66 x $10^{-7}$ m$^3$K/kg) than does the stoichiometric sample (3.8 x $10^{-7}$ m$^3$K/kg). Neglecting the very small $\theta$ and putting the molecular weight $M$ into Eq. (2) we can calculate the effective localized magnetic moment $\mu_{eff}$ at low temperature for the two samples. The values $C$ obtained from the data fit in Fig. 4(a) yield effective moments $\mu_{eff}$ for NaRhO$_2$ of 2.0 x $10^{-1}$ $\mu_B$ and for Na$_{0.3}$RhO$_2$·0.6H$_2$O of 0.8 x $10^{-1}$ $\mu_B$ per formula unit. This is surprising, as the oxidation of Rh$^{3+}$ to Rh$^{4+}$ should lead to a higher effective moment. Both samples are paramagnetic at room temperature, Fig. 4(b); however, Na$_{0.3}$RhO$_2$·0.6H$_2$O shows a slight ferromagnetic feature (saturation magnetization $\sigma_S = 2 \times 10^{-3}$ Am$^2$/kg (2 x $10^{-3}$ emu/g)) together with the paramagnetic behavior (Fig. 4(b) inset). At the present time this room-temperature ferromagnetism is attributed to inhomogeneous areas in the layered material where the lattice is locally distorted to meet the requirements for ferromagnetism. This conclusion is corroborated by the EXAFS measurements, which indicate that significantly more disorder is present in the Na$_{0.3}$RhO$_2$·0.6H$_2$O sample. The possibility of contaminating ferromagnetic elements present in the system, not detectable within the limit of our experimental techniques, does remain. For example, the magnitude of the measured room-temperature ferromagnetic moment corresponds to a possible Fe impurity on the order of $10^{-3}$ wt%. However, this possibility is doubtful as numerous investigations revealed the ferromagnetic signal to be present *only* in the Na-deficient sample and not in the stoichiometric sample, despite repeated synthesis procedures.

In conclusion we have synthesized the first rhodium monolayer oxyhydrate Na$_{0.3}$RhO$_2$(H$_2$O)$_{0.6}$ which exhibits simultaneously a semiconducting behavior and a weak ferromagnetic component at room temperature. At the current time, the origin of the weak ferromagnetic component remains elusive but is tentatively attributed to structural disorder.

Acknowledgements: Research at Brookhaven National Laboratory is supported by the U. S. Department of Energy, Office of Basic Energy Sciences, Division of Materials Science and the Division of Chemical Sciences under contract DE-AC02-98CH10886. Research carried out in part at the NSLS at BNL is supported by the U.S. DOE (DE-Ac02-98CH10886 for beam line X7A).

# Figure captions :

**Fig. 1:**
Rietveld refinement fit of the structural model of $Na_{0.3}RhO_2 \cdot 0.6H_2O$ to room-temperature synchrotron X-ray powder diffraction data. The tick marks below the data indicate the positions of the allowed reflections. The lower curves represent the differences between observed and calculated profiles ($I_{obs}$ - $I_{calc}$) plotted on the same scale as the observed data. Data near 10° 2θ are due to a small impurity and are excluded. [(Inset) Polyhedral representation of the structure of $Na_{0.3}RhO2 \cdot 0.6H_2O$ viewed along the *a*-axis (left). For comparison, the structure of $NaRhO_2$ is shown in the center figure. Stacking schemes of the respective $RhO_2$ layers are emphasized in the diagrams at right. Dotted lines in the figures define unit cells.]

**Fig. 2:**
Resistivity as a function of temperature for $NaRhO_2$ and $Na_{0.3}RhO_2 \cdot 0.6H_2O$. The circles and arrows emphasize the distinct scales for resistivity.

**Fig. 3:**
(a) The normalized Rh K-edge XANES spectra for $NaRhO_2$ and $Na_{0.3}RhO_2 \cdot 0.6H_2O$. (b) Fourier-transform magnitudes of the Rh K-edge EXAFS spectra for $NaRhO_2$ and $Na_{0.3}RhO_2 \cdot 0.6H_2O$. (The Fourier transforms are not phase-corrected, so the indicated distances differ from the crystallographic structure.)

**Fig. 4:**
(a) Temperature dependence of the experimental and estimated dc susceptibilities for $Na_{0.3}RhO_2 \cdot 0.6H_2O$ and $NaRhO_2$ at 10 kOe in a temperature range 5 to 50 K. The solid lines indicate the fit to the data utilizing eq. 2. (b) Magnetization (M) as a function of applied magnetic field (H) at 300 K for $NaRhO_2$ and $Na_{0.3}RhO_2 \cdot 0.6H_2O$. The inset shows temperature dependence of the experimental susceptibilities for $Na_{0.3}RhO_2 \cdot 0.6H_2O$ and $NaRhO_2$ at 10 kOe in a temperature range 5 to 300 K.

Fig. 1

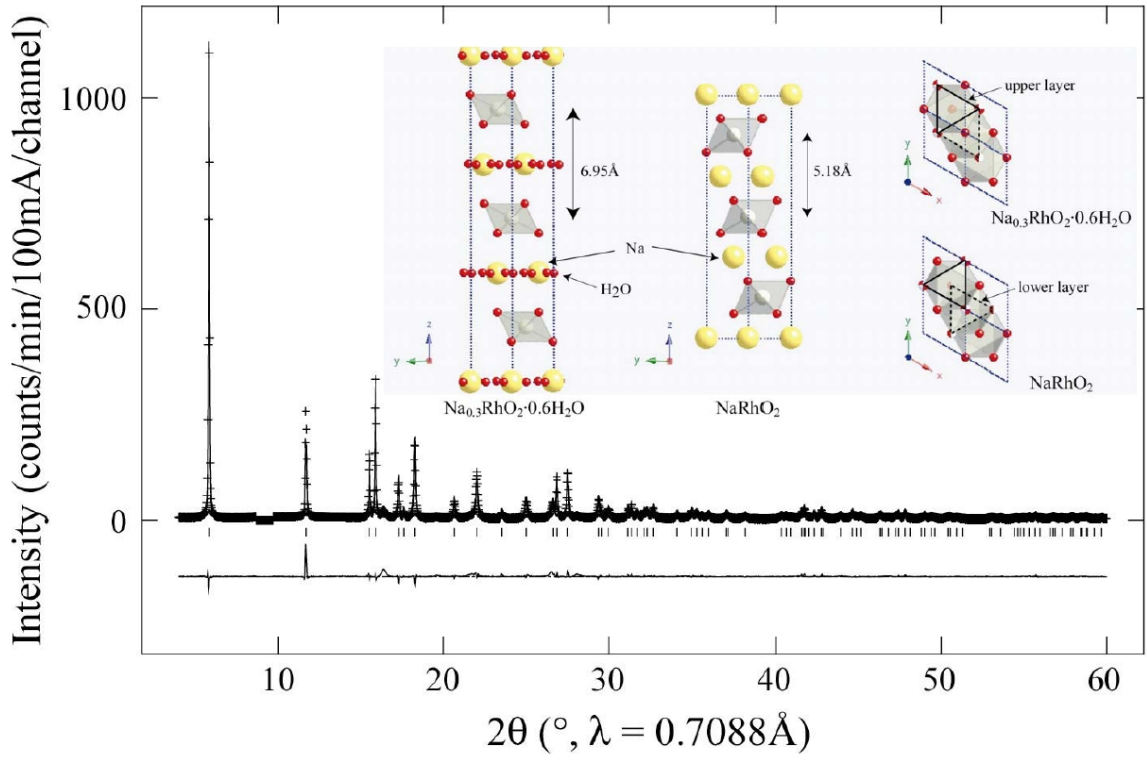

Fig. 2

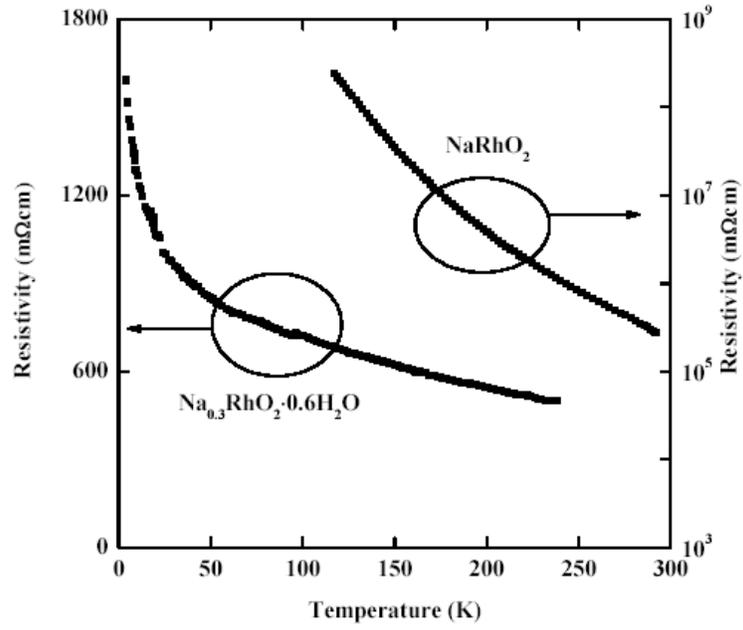

Fig. 3

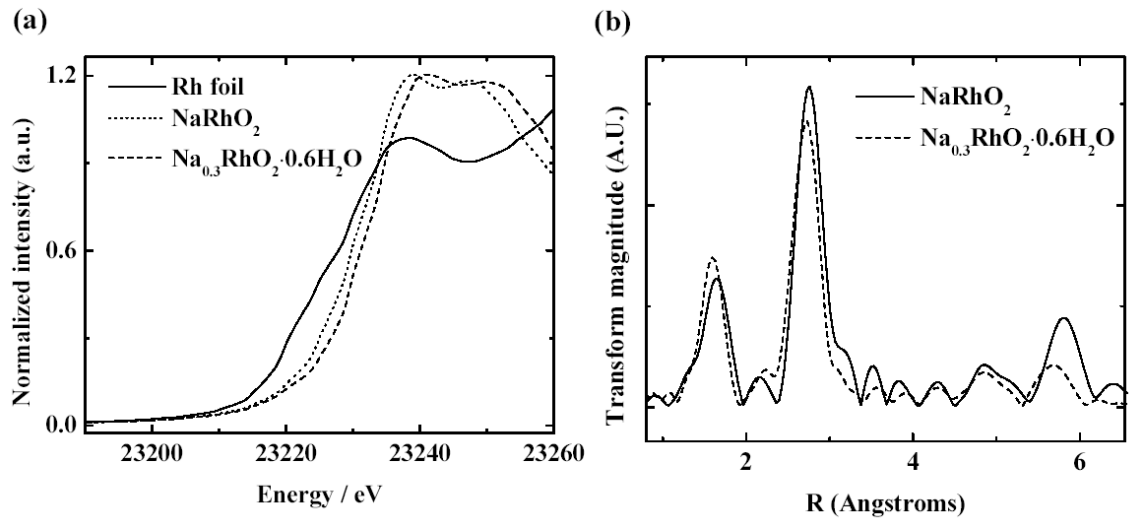

Fig. 4

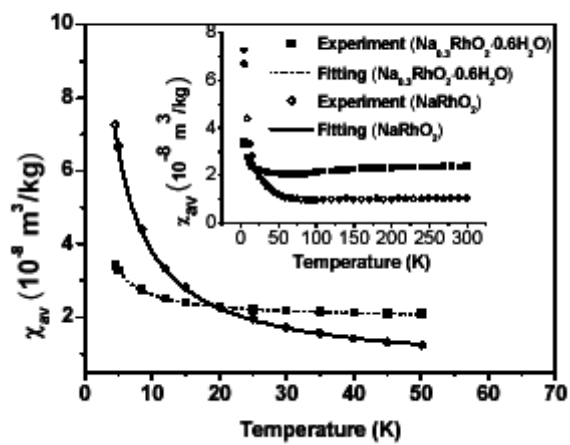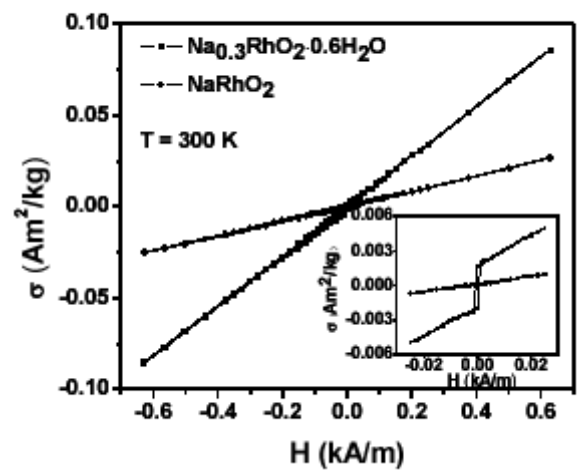

Table 1. Atomic coordinates of $Na_{0.3}RhO_2 \cdot 0.6H_2O$.

| Atom | x | Y | z | occupancy | $U_{iso}$ (Å$^2$) |
|---|---|---|---|---|---|
| Rh | 0 | 0 | 1/2 | 1 | 0.006(1) |
| O | 0 | 0 | 0.1213(1) | 1 | 0.002(1) |
| Na | 0 | 0 | 0 | 0.3 | 0.029(4) |
| OW | 0.182(3) | 0.365(7) | 0.001(2) | 0.1 | 0.029(4) |

Space group ***R*-3m**; $a$ = 3.0542(1) Å, $c$ = 20.8560(9) Å, $V$ = 168.48(1) Å$^3$. $_wR_p$ = 11.7%, $R_p$ = 7.72%. OW denotes water oxygen atom. Constraints were used to make the isotropic displacement parameters the same for interlayer atoms. Bond distance constraints were used to fix Rh-O = 2.015(1) Å.

Table 2. Selected bond distances of $Na_{0.3}RhO_2 \cdot 0.6H_2O$.

| Rh-O* | 2.002(1) × 6 | OW-O | 2.69(4) |
|---|---|---|---|
|  |  |  | 2.72(4) |
| Na-O | 2.529(3) × 2 | OW-OW | 2.649(2) × 4 |
| Na-OW | 2.27(1) × 6 |  |  |

*Bond distance constraints were used.